\documentstyle[12pt]{article}

 1
\font\elevenrm=cmr10 scaled\magstep 1
 1

\begin{document}
\def\psfig#1{}

\newcommand{\beq}{\begin{eqnarray}}
\newcommand{\eeq}{\end{eqnarray}}
\newcommand{\nn}{\noindent}
\newcommand{\non}{\nonumber}
\newcommand{\ee}{\mbox{$e^+ e^-$}}
\newcommand{\ra}{\rightarrow}
\newcommand{\s}{\\ \vspace*{-3mm} }

\nn \hspace*{11cm} UdeM-GPP-TH-95-18 \\
\vspace*{0.3cm}

\centerline{\large{\bf SEARCHING FOR NEW MATTER PARTICLES}}

\vspace*{1.5mm}

\centerline{\large{\bf AT  FUTURE COLLIDERS\footnote{Talk given at ``Beyond
the Standard Model IV", Lake Tahoe, Ca., December 1994.}}}

\vspace*{3mm}

\centerline{\sc A.~DJOUADI}

\vspace*{1mm}

\centerline{\it Groupe de Physique des Particules, Universit\'e de
Montr\'eal}
\centerline{\it Case 6128 A, H3C 3J7  Montr\'eal PQ, Canada.}

\vspace*{0.01cm}

\begin{small}
\begin{center}
\parbox{14cm}
{\begin{center} ABSTRACT \end{center}

\nn We discuss the production of new particles that are predicted by many
extensions of the Standard Model, at future high--energy pp, eP and e$^+$e$^-$
linear colliders. We focus on the case of exotic, excited and di--fermions.}
\end{center}
\end{small}

\vspace*{0.01cm}

\subsection*{1.~Introduction}

Many theories beyond the Standard Model (SM) of the electroweak and strong
interactions, such as Grand Unified Theories, Composite Models or Technicolor,
predict  the existence of new matter particles. These particles can be cast
into three categories:  exotic fermions, excited fermions and difermions. In
this talk, I will discuss the prospects for producing some of these new
particles at future $pp$ [LHC with $\sqrt{s}= 14$ TeV], $eP$ [LEP$\times$LHC
with $\sqrt{s}=1.2$ TeV] and $\ee$ linear colliders [NLC with $\sqrt{s}=0.5$--1
TeV with its $\ee, \gamma \gamma$ and $e \gamma$ modes]. The present write--up
summarizes a study which has been performed in a recent  report [1], to which
we
refer for a more detailed discussion and for a  complete  set of references.

\subsection*{2.~Exotic Fermions}

These fermions are exotic with respect to their transformation  under the SM
group: they have the usual lepton and baryon quantum numbers but non--canonical
SU(2)$_L \times$ U(1)$_Y$  quantum numbers, e.g. the left--handed components
are
in weak isosinglets  and/or  the right--handed components in weak isodoublets.
These particles are predicted by Grand Unified Models which have a single
representation into which a  complete generation of SM quarks and leptons can
be
naturally embedded. In most cases, these representations are large enough to
accommodate new fermions which, in fact, are needed to have anomaly--free
theories. For  instance, in the group E$_6$ which is suggested as a low energy
limit of some superstrings theories, each fermion generation  lies in the {\bf
27} representation and, in addition to the 15 SM  chiral fields, 12 new fields
are needed to complete the representation.

The classic examples of new fermions include: sequential 4th generation
fermions [with the neutrino having a right-handed component for its mass to
be generated in a gauge invariant way]--such fermions are not truly `exotic'
in the usual sense; vector fermions with both left- and right-handed components
in weak isodoublets [in E$_6$ there are two--isodoublets of heavy leptons for
each generation, and an isosinglet vector colored particle], mirror fermions
which have the opposite chiral properties as the SM fermions and isosinglet
fermions [which are more frequently discussed in the literature] such as the
SO(10) Majorana neutrino.

It is conceivable that these fermions, if for instance they are protected by
some symmetry, acquire masses not much larger than the Fermi scale. This is
even
necessary for purposes of anomaly cancellation if the new gauge bosons which
are generic predictions of Grand Unified Theories  are relatively light. In the
case of  sequential and mirror fermions [at least in the simplest models where
the SSB pattern is the same as in the SM], unitarity arguments suggest that the
masses should not exceed a few hundred GeV. These particles, if they exist,
could be  therefore accessible at the next generation  of colliders.

Except for singlet neutrinos, the new fermions couple to the photon and/or to
the electroweak gauge bosons $W/Z$ [and for heavy quarks, to gluons] with full
strength; these  couplings allow for pair production with practically
unambiguous cross sections. If they have  non--conventional quantum numbers,
the
new fermions will mix with their SM  partners. This mixing will give rise to
new
currents which determine the decay  properties of the heavy fermions and allow
for their single production. If the  mixing between different generations
[which
induces FCNC at tree--level] is  neglected, the mixing pattern simplifies. The
remaining angles are restricted by LEP and low energy experiment data to be
smaller than ${\cal O}(0.05-0.1)$. Note that LEP1 sets bounds of order $\sim
M_Z/2$ on the masses of  these particles [stronger mass bounds from Tevatron
can
be set for  quarks]; masses up to $M_Z$ might be probed at LEP2.

The heavy fermions decay through mixing into massive gauge bosons plus their
ordinary light partners, $F \ra fZ/ f'W$. For masses larger than $M_W(M_Z)$ the
vector bosons will be on--shell. For small mixing angles, $\zeta<0.1$, the
decay widths are less than 10 MeV (GeV) for $m_F= 0.1 (1)$ TeV. The  charged
current decay mode is always dominant and for $m_F \gg M_Z$, it has a
branching fraction of 2/3.

If their masses are smaller than the beam energy, the new fermions can be pair
produced in $\ee$ collisions, $\ee \ra F\bar{F}$, through $s$--channel  gauge
boson exchange. The cross sections are of the order of the point--like  QED
cross section and therefore,  are rather large; Fig.~1a. Because of their clear
signatures, the detection of these  particles is straightforward in the clean
environment of $\ee$ colliders, and masses very close to the kinematical limit
can be probed. The total cross sections, angular  distributions and the
polarization of the final particles allow one to  discriminate the different
types of fermions.  Charged fermions can also be  pair--produced at
$\gamma\gamma$  colliders, $\gamma \gamma \ra F\bar{F}$, and for relatively
small masses, the  cross sections can be larger than in the  $\ee$ mode.

Heavy quarks can be best searched for at hadron colliders where the production
processes, $gg/ q \bar{q} \ra Q \bar{Q}$, give very large cross sections: at
LHC
with $\sqrt{s}=14$ TeV and a luminosity of 10 fb$^{-1}$ quark masses up to 1
TeV
can be reached using the spectacular $ZZqq \ra 4l^\pm qq$ signal [with a
branching  ratio of ${\cal O} (10^{-3})$]; see Fig.~2a.

\vspace*{-2mm}

\begin{figure}[htbp]
\hspace*{-0.5cm}
\centerline{\psfig{figure=tpair.ps,height=8.5cm,width=7cm}
\hspace*{0.2cm}
            \psfig{figure=tsing.ps,height=8.5cm,width=7cm}}
\vspace*{-.5cm}
\caption{\small Total cross sections for the (a) pair and (b) single (first
generation) production of exotic leptons at the NLC with $\protect \sqrt{s}=1$
TeV.}
\end{figure}

\vspace*{-8mm}

\begin{figure}[htbp]
\centerline{\psfig{figure=fig7.ps,height=8.5cm,width=7.5cm,angle=90}
\hspace*{-.5cm}
           \psfig{figure=fig6.ps,height=8.5cm,width=7.5cm,angle=-90}}
\vspace*{-.5cm}
\caption{\small Total cross sections for the pair production of heavy quarks
at LHC and for the single production of exotic leptons at LEP$\times$LHC.}
\end{figure}

\bigskip

For not too small mixing angles, one can also have access to the new fermions
via single production in association with their light partners. The rate for
this type of process is much more model dependent but can substantially
increase the reach of a given accelerator. In $\ee$ collisions, this proceeds
only via $s$--channel $Z$ exchange  in the case of  quarks and second/third
generation leptons, leading to small rates. For the first generation leptons,
however, one has additional $t$--channel exchanges which increase the cross
sections by several orders of magnitude; see Fig.~1b.

A full simulation  of the signals [$\ee \ra \nu_e e jj$ for $N$ and  $\ee \ra
e^+e^- jj$ for $E$] and the backgrounds [mainly from single and pair
production of $W/Z$ bosons] has been performed  using a model detector,
assuming
$\sqrt{s}=500$  GeV and $\int {\cal L}=50$ fb$^{-1}$.  For mixing angles
$\zeta=0.05$ for $E$ and $\zeta=0.025$ for $N$,  it has been shown that one
can reach lepton masses up to $m_L \sim 450$ GeV. For $m_L=350$ GeV, smaller
values of the mixing angles ($\zeta  \sim 0.005$ for $N$ and  $\zeta \sim
0.03$ for $E$) can be probed. The angular distributions and  the final
polarization allow to discriminate between fermions with left- and right-handed
mixing, and between neutrinos of  Dirac  or Majorana type.

Heavy leptons of the first generation can also be singly produced in $eP$
collisions through $t$--channel exchange; the cross sections are shown in
Fig.~2b. At LEP$\times$LHC with $\sqrt{s} =1.2$ TeV and $\int {\cal L}=2$
fb$^{-1}$, masses up to  700 GeV for $N$ and 550 GeV for $E$ can be probed, for
$\zeta \sim 0.1$.  Note that heavy neutrinos of  Left--Right models can be
produced in $eP$ collisions  via $t$-channel $W_R$  exchange; one finds that at
LEP$\times $LHC, the discovery reach in the $m_N- M_{W_R}$ plane is
$m_N+0.38M_{W_R}<1090$  GeV. In $\ee$ collisions these  neutrinos, when
kinematically allowed can be  pair produced if the right-handed $W$ bosons are
not too heavy [$M_{W_R}< 2$ TeV at $\sqrt{s}=0.5$ TeV].

\subsection*{3.~Excited Fermions}

The existence of excited particles is a characteristic signal of substructure
in the fermionic sector: if the known fermions are composite, they  should be
the ground state of a rich spectrum of excited states which decay down to the
former states via a magnetic dipole type de-excitation. However,  a
satisfactory
dynamical model is still lacking and a phenomenological input is needed to
study
this possibility. For simplicity, it is assumed that the  excited fermions have
spin and isospin 1/2; their couplings to gauge bosons are vector--like [form
factors and new contact interactions may also be present] and that the coupling
which describes the transition between excited and ordinary fermions is chiral
and inversely proportional to the compositeness scale $\Lambda$ which is of
${\cal O}$(1 TeV).

The excited fermions decay into gauge bosons and their ordinary partners,
$f^\star \ra fV$. The charged leptons have the electromagnetic decay which
has at least a branching ratio of 30\%, and the excited quarks decay most of
the time into quarks and gluons. These two decays constitute a very
characteristic signature and discriminate them from the exotic fermions
previously discussed.

If kinematically allowed, excited fermions can be pair-produced without any
suppression due to powers of $1/\Lambda$.  In
$e^{+}e^{-}$  and $\gamma \gamma$ collisions, the processes and the cross
sections are the same as for vector exotic fermions; the only difference is in
the  decay modes. Since the latter can be easily searched for, $f^*$ with
masses near $\sqrt{s}/2$ can be discovered in these machines.
Excited fermions can be singly produced with their light partners and  the
rates
are suppressed by the factor $1/\Lambda^2$. At $\ee$ colliders, for  quarks and
second/third generation leptons, for which the process is mediated by
$s$--channel boson exchange, the cross sections are very small. But for the
first generation excited fermions, one has substantial  contributions due to
additional $t$--channel diagrams: $W$ exchange for  $\nu_e^*$ and $Z/\gamma$
exchanges for $e^*$, which increase the cross sections by several orders of
magnitude. At a 1 TeV $\ee$ collider, the cross section for $e^*$ is larger
than
1 pb across the entire mass range for $\Lambda=1$ TeV, while for $\nu_e^*$ it
drops to 100 fb for $m_{\nu^*} \sim 900$ GeV; Fig.~3a. $e^*$ and $\nu_e^*$ can
be also singly produced in $e\gamma$ collisions [the former as a resonance and
the latter in association with a $W$] with much larger rates, and all excited
fermions can be singly produced in $\gamma \gamma$ collisions with appreciable
cross sections. For $\Lambda \sim $ few TeV, and  if  kinematically  allowed,
excited quarks and  leptons can therefore easily be found in such machines.

Due to the special couplings of the electron to the excited leptons of the
first generation, one can have single production of $e^*$ through $t$-channel
$\gamma$ and $Z$ exchange, and $\nu_e^*$ through $W$  exchange in $eP$
collisions. For $e^*$ production, one has three different contributions  and
the
cross section is an order of magnitude larger than for $\nu_e^*$.  Requiring a
few tens of  events, masses up to 800 GeV for $\nu^*$ and $e^*$ can be probed
at
LEP$ \times$LHC for $\int {\cal L}=2$ fb$^{-1}$. Background problems make that
the detection of the first generation excited quarks is more difficult.

\vspace*{2mm}

\begin{minipage}[t]{7.5cm}
\psfig{figure=xsing.ps,height=8cm,width=6.5cm}
\end{minipage}
\hspace*{2mm}
\begin{minipage}[t]{7.5cm}
\psfig{figure=fig11a.ps,height=8cm,width=6.5cm,angle=-90}
\end{minipage}
\begin{figure}[htbp]
\vspace*{-5mm}
\caption{\small Total production cross sections for single excited leptons
at NLC (a) and for single excited quarks (and for the QCD background) at LHC
(b).}
\end{figure}

Excited quarks can be produced in $pp$ collisions through a variety of
mechanisms. The dominant production channel is the gluonic excitation of quarks
$g + q \ra q^{*}$ which occurs through the $q^*qg$ magnetic interaction;
the signature is dijet mass bumps.  [Through preon interactions excited
quarks and also leptons, could eventually  be produced at observable rates.]
The cross section are large, and the QCD backgrounds have been shown to be
under control; Fig.~3b.   At LHC with $\sqrt{s}=14$ TeV and a luminosity of 10
fb$^{-1}$, based on 100  to 1000 events, a mass range of 5--6 TeV can be
reached
for excited quarks.

\subsection*{3.~Difermions}

Difermions are scalar or vector particles [spin 1/2 difermions are also
discussed  in the context of supersymmetric theories] which have unusual baryon
and/or  lepton quantum numbers. Examples of these  particles are leptoquarks
(LQ) with  B$=\pm 1/3$ and L$=\pm 1$, diquarks with B$=\pm 2/3$ and L$=0$ and
dileptons  with B$=0$ and L$=\pm2$. They are predicted in GUT's [e.g., in
E$_6$, the color triplet weak isosinglet particle in the {\bf 27}-plet can  be
either a LQ or diquark] and in composite models.

In addition to the usual couplings to gauge bosons, difermions have couplings
to fermion pairs which determine their decays [here also one can neglect the
couplings between different generations to prevent FCNC at tree-level].  These
couplings are a priori unknown.  In the case of LQ's for example, a systematic
description of their quantum numbers and interactions can be made by starting
from an effective lagrangian with general SU(3)$\times$SU(2)$\times$U(1)
invariant couplings and conserved  B and L numbers. This leads to the existence
of 5 scalar and 5 vector LQ's  with distinct SM transformation properties. In
general, present data constrain  difermions to have masses larger than 50--150
GeV.

Leptoquarks can be produced in pairs at $\ee$ colliders through gauge boson
exchange; significant $t$-channel quark exchange can be present in some
channels if the quark-lepton-LQ couplings are not too small. Depending on  the
charge, the spin and isospin of the LQ, the cross sections can vary  widely. As
an example, at $\sqrt{s}=500$ GeV and assuming $m_{LQ}\sim 200$  GeV, $\sigma$
varies between 7 fb  [for the scalar isosinglet with charge $-1/3 $] and 3.3 pb
[for the vector iso--triplet].  Through the signatures of 2 leptons plus 2
jets,
these states are accessible for masses smaller  than the beam energy.  The
study
of the various final states and the angular  distributions would allow the
determination of the quantum number of the LQ's as in the case of exotic
fermions. LQ's can also be  pair produced in $\gamma \gamma$ collisions, just
as
were the exotic fermions; depending on the LQ charge, the cross sections can
be much larger or much smaller than for charged leptons.

Single production of scalar and vector leptoquarks can also take place in the
$\ee$, $e\gamma$ and $\gamma \gamma$ modes of the collider. The kinematical
reach is thus extended to $\sqrt{s}$ but the production rates are suppressed
by
the unknown LQ coupling to quark--lepton pairs,  $\sqrt{k \alpha}$. The cross
sections are shown in Fig.~4. The most important subprocess in this case is
$e\gamma  \ra q$+LQ and the contributions of both direct and resolved photons
have to be taken into account. At a 1 TeV $\ee$ collider with $\int{\cal L}=60$
fb$^{-1}$,  one reaches masses close to $\sqrt{s}$ [i.e. 1 TeV for $\ee$,
$\simeq 0.9$ TeV  for $e \gamma$ and $\simeq 0.8$ TeV for $\gamma \gamma$). At
$e \gamma$  colliders, first generation LQ's are observable for $k$ as
small as  $10^{-3}$.

\vspace*{-7mm}

\begin{figure}[htbp]
\hspace*{-.1cm}
\centerline{\psfig{figure=eeeggg.ps,height=8.5cm}}
\vspace*{-.9cm}
\caption{\small Cross sections for the single production of scalar leptoquarks
at $\ee, e\gamma$ and $\gamma \gamma$ colliders: (a) $Q=-5/3$ or $-1/3$ and
(b) $Q=-4/3$ or $-2/3$. }
\end{figure}

\vspace*{-9mm}

\begin{figure}[htbp]
\hspace*{-1cm}
\centerline{\psfig{figure=vlq2a.ps,height=8.5cm,width=7.5cm,angle=90}
\hspace*{-.5cm}
            \psfig{figure=vlq4b.ps,height=8.5cm,width=7.5cm,angle=90}}
\vspace*{-1.1cm}
\caption{\small (a) Total cross sections for pair production of scalar
(dash-dot) and vector LQ's (the dot, dash, dash-dot lines correspond to the
$qq, gg$ and total cross sections) at LHC. (b) Total cross sections for single
production of vector LQ's at LHC: the dot(dash-dot) is for the $gu(gd)$
subprocess and for $\kappa=1,0$. }
\end{figure}

\bigskip

Since they are strongly interacting particles, LQ's can be produced at  hadron
colliders with very large rates. At LHC, pair production in the $gg/q  \bar{q}
\ra$ LQ+LQ process leads to cross sections ranging from a few nb for masses
around 100 GeV to a few fb for masses ${\cal O}$(1 TeV). The rate for vector
particles [which depends on their anomalous magnetic  moment $\kappa$] is
substantially higher than for scalar particles; Fig.~5a. At LHC with 100
fb$^{-1}$  the search reach for scalar/vector LQ's is 1.4/2.2(1.8) TeV for
$\kappa=1(0)$, if one assumes a branching fraction of unity for the $eejj$
final
state.

Single scalar LQ production, through $gu\to e^+$LQ and $gd\to \bar \nu$LQ, can
also lead to large cross sections for Yukawa couplings $k ={\cal O}(1)$; in
this
case masses up to $\sim$ 1.5 TeV can be reached at the LHC.  Vector leptoquarks
have larger production rates and the discovery reach can be  extended to $\sim$
2 TeV; Fig.~5b.

Note that $eP$ colliders are ideal machines for the production of the first
generation leptoquarks.  The latter  can be produced as $s$ channel resonances
in $e \ra$LQ, unless the coupling $k$ is extremely tiny. At LEP$\times$LHC
masses up to 1 TeV can be reached, even for $k$ values a few orders of
magnitude
smaller than unity.

Dilepton production has been considered at $\ee$ colliders in the three modes
$\ee, e\gamma$ and $\gamma \gamma$. These particles are  accessible up to
masses
close to $\sqrt{s}/2$ in pair production, $\ee/\gamma  \gamma \ra
X^{++}X^{--}$:
the rates [especially in $\gamma \gamma$ collisions because of the charge]  are
very large and the signatures [four leptons] are  spectacular. Dileptons can
also be singly produced in the three modes of the collider. In particular, at
1
TeV $\ee$ collider, scalar and vector dileptons  can be observed up to masses
of
$\sim 0.9$ TeV in the $e\gamma$ mode  even for couplings to lepton  pairs as
small as $10^{-3}$ the electromagnetic coupling; in the $\ee$ and  $\gamma
\gamma$ modes, dileptons can be observed for couplings an order of  magnitude
larger.

Finally, diquarks can be pair produced in $\ee$ and $\gamma \gamma$ collisions
for masses smaller than $\sqrt{s}/2$ with appreciable rates, with a signal
consisting of an excess of 4 jets events. They can be also pair produced at
hadron colliders, either in pairs or singly [for the first generation]  if the
couplings to quark pairs is not too small. However, since the signals consist
only in jets, the large QCD backgrounds might be a problem.  In the  case of
single production, as a dijet resonance, the detection of these  particles
depends on excellent energy resolution being available.

\bigskip

\nn {\bf Acknowledgements:} I thank H.E. Haber, J. Ng, T.G. Rizzo and the
members of the ``New Particles and Interactions" subgroup in Ref.~[1] for
discussions.

\subsection*{References}

1. A. Djouadi, J. Ng and T. G. Rizzo [conv.] et al., to appear in {\it
Electroweak Symmetry Breaking and Beyond the Standard Model}, T. Barklow, S.
Dawson, H.E. Haber and T. Siegrist, eds, World Scientific.

\end{document}